\documentstyle[twoside,titlepage,psfig]{article}

\def\pp{\par\parshape 2 0truecm 16.0truecm 1truecm
  15.0truecm\noindent}

\begin{document}
\font\utap=cmmib10 scaled\magstep2
\font\urc=cmmib10 scaled\magstephalf
\font\fontname=cmcsc10 
\newcommand{\lbkt}[1]{\left[#1\right]}
\newcommand{\mbkt}[1]{\left\{#1\right\}}
\newcommand{\sbkt}[1]{\left(#1\right)}
\newcommand{\ngt}{\raisebox{-0.8ex}{~$\stackrel{\textstyle >}{\sim}$}~}
\newcommand{\nlt}{\raisebox{-0.8ex}{~$\stackrel{\textstyle<}{\sim}$}~}
\newcommand{\msun}{\,\mbox{M}_{\odot}}
\newcommand{\abs}[1]{\left|#1\right|}
\newcommand{\vect}[1]{\mbox{\boldmath $#1$}}
\newcommand{\gtsima}{$\; \buildrel > \over \sim \;$}
\newcommand{\ltsima}{$\; \buildrel < \over \sim \;$}
\newcommand{\simgt}{\lower.5ex\hbox{\gtsima}}
\newcommand{\simlt}{\lower.5ex\hbox{\ltsima}}
\newcommand{\himpc}{{\hbox {$h^{-1}$}{\rm Mpc}} }
\newcommand{\PS}{ {\scriptscriptstyle {\rm PS}} }
\newcommand{\M}{ {\scriptscriptstyle {\rm M}} }
\newcommand{\0}{ {\scriptscriptstyle {0}} }
\renewcommand{\thesection}{\normalsize\bf\arabic{section}}
\renewcommand{\thesubsection}
{\normalsize\it\arabic{section}.\normalsize\it\arabic{subsection}. }
\renewcommand{\theequation}{\mbox{\rm
    {\arabic{section}.\arabic{equation}}}}
\pagestyle{myheadings} \markboth{\fontname statistical properties of
  X-ray clusters} {\fontname Kitayama~~\&~~Suto}
\begin{titlepage}
\vspace*{-1.5cm}
\begin{minipage}[c]{3cm}
  \psfig{figure=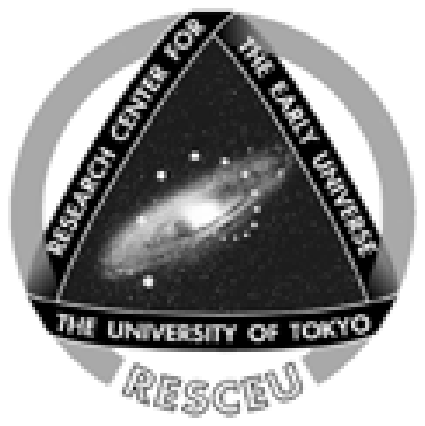,height=3cm}
\end{minipage}
\begin{minipage}[c]{10cm}
\begin{centering}
{
\vskip 0.1in
{\large \sf 
THE UNIVERSITY OF TOKYO\\
\vskip 0.1in
Research Center for the Early Universe}\\
}
\end{centering}
\end{minipage}
\begin{minipage}[c]{3cm}
\vspace{2.5cm}
RESCEU-10/96\\
UTAP-229/96
\end{minipage}\\
\vspace{1cm}

\addtocounter{footnote}{1}
 
\baselineskip=20pt

\begin{center}

  {\Large\bf Semi-analytic predictions for statistical properties of
    X-ray clusters of galaxies in cold dark matter universes}

\vspace{1cm}
 
{\sc Tetsu Kitayama$^{1}$ and Yasushi Suto$^{1,2}$}
\end{center}

\baselineskip=16pt 
\noindent {\it $^{1}$ Department of Physics, School
of Science, The University of Tokyo, 
Tokyo 113, Japan.}

\noindent {\it $^{2}$ Research Center For the Early Universe (RESCEU), 
School of Science, The University of Tokyo, Tokyo 113, Japan.}

\centerline { e-mail: kitayama@utaphp1.phys.s.u-tokyo.ac.jp, ~
suto@phys.s.u-tokyo.ac.jp}

\vspace{8mm}

\centerline{Received 1996 March 13;  accepted 1996 April 16}

\vspace{1.2cm}

\baselineskip=14pt

\centerline {\bf Abstract}

\noindent
Temperature and luminosity functions of X-ray clusters are computed
semi-analytically, combining a simple model for the cluster gas
properties with the distribution functions of halo formation epochs
proposed by Lacey \& Cole (1993) and Kitayama \& Suto (1996). In
contrast to several previous approaches which apply the
Press--Schechter mass function in a straightforward manner, our method
can explicitly take into account the temperature and luminosity
evolution of clusters.  In order to make quantitative predictions in a
specific cosmological context, we adopt cold dark matter (CDM)
universes.

Assuming the baryon density parameter $\Omega_{\rm B}=0.0125 h^{-2}$
($h$ is the Hubble constant in units of
100km$\cdot$sec$^{-1}\cdot$Mpc$^{-1}$) and the {\it COBE}
normalization of matter fluctuations, temperature and luminosity
functions of X-ray clusters depend sensitively on the density
parameter $\Omega_0$.  Allowing for several uncertainties in
observational data as well as in our simplified assumptions, we
conclude that $\Omega_0 \sim 0.2-0.5$ and $h\sim 0.7$ CDM models
with/without the cosmological constant reproduce simultaneously the
observed temperature and luminosity functions of X-ray clusters at
redshift $z\sim0$.

\medskip

\noindent {\it Subject headings:} cosmology: theory -- dark matter --
galaxies: clusters -- galaxies: formation -- X-rays: galaxies
\vfill

\centerline{\sl The Astrophysical Journal (1996), October 1 issue, in
  press.} \eject

\end{titlepage}

\baselineskip=14pt
\setcounter{page}{2}

\section{\normalsize\bf Introduction}
\setcounter{equation}{0}

Since clusters of galaxies are the largest virialized structure in the
universe, they should retain the initial conditions at their formation
epochs fairly faithfully. This implies that detailed studies of the
clusters at high redshifts, as well as at present, should provide
important clues to the evolution of the universe itself.  Since X-ray
identifications of clusters of galaxies are largely free from the
projection effect which notoriously plagues optically selected cluster
catalogues, X-ray observations are suitable for probing cosmological
signatures from clusters of galaxies. Homogeneous samples of distant
clusters of galaxies, which recent X-ray satellites such as {\it
  ROSAT} and {\it ASCA} are constructing, will uncover various
statistical properties of clusters with higher reliability in the near
future. Therefore, quantitative theoretical predictions are of great
value in interpreting the observed data properly.

Most theoretical approaches in X-ray cosmology rely on either
state-of-the-art numerical simulations, or simplified analytical
formalisms. The former approach is limited by the dynamical range
available on the present computer resources; a typical core radius of
clusters is ($0.1\sim 0.2 \himpc$) while the mean separation of the
Abell clusters (richness class 1) is $\sim 55\himpc$, where $h$ is the
Hubble constant $H_0$ in units of $100
\mbox{km}\cdot\mbox{sec}^{-1}\cdot\mbox{Mpc}^{-1}$. A small-scale
resolution much below the core size is essential because a large
fraction of the X-ray luminosity of clusters comes from the core. On
the other hand, a large simulation box size is a prerequisite for
statistical studies of clusters. Unfortunately, it is still hard to
simultaneously satisfy these requirements even with the currently most
advanced simulations (e.g. Kang et al. 1994; Bryan et al. 1994; Cen
et al. 1995).

A major fraction of the X-ray luminosity from clusters is produced via
a fairly simple process, thermal bremsstrahlung. Thus one may readily
compute their temperature and luminosity functions at redshift $z$,
$n_{\rm T}(T,z)$ and $n_{\rm L}(L,z)$, once the mass function $n_{\rm
  M}(M,z)$ is given, where $T$, $L$ and $M$ are the temperature,
luminosity and mass of the clusters, respectively. Although the
Press--Schechter theory (Press \& Schechter 1974, hereafter PS) is
frequently applied for this purpose, it has a serious limitation in
predicting the temperature and luminosity functions; PS theory
predicts the number density of virialized objects of mass $M$
collapsed {\it before} a given epoch $z$, but does not specify the
formation epoch $z_{\rm f}$ of the objects. In fact, the predictions
of the spherical nonlinear collapse model (e.g. Peebles 1980) suggest
that the temperature and luminosity of objects that virialize at
$z_{\rm f}$ should scale as $T(z_{\rm f}) \propto (1+z_{\rm f})$ and
$L(z_{\rm f}) \propto (1+z_{\rm f})^{7/2}$ in the Einstein--de Sitter
universe, for instance. The previous approaches based on the PS
formula (e.g. Evrard \& Henry 1991; Hanami 1993) have simply replaced
$z_{\rm f}$ by $z$ in computing $T$ and $L$ (see also
eq.~[\ref{tempfnps}] below). This procedure corresponds to implicitly
assuming that $T$ and $L$ of individual clusters {\it decline} with
time as $T(z)/T(z_{\rm f})= (1+z)/(1+z_{\rm f})$ and $L(z)/L(z_{\rm
  f})= (1+z)^{7/2}/(1+z_{\rm f})^{7/2}$.  Numerical simulations, on
the contrary, suggest modest evolution in the opposite direction
(e.g. Evrard 1990; Suginohara 1994; Navarro, Frenk, \& White 1995).
While the above assumptions may not alter $n_{\rm T}(T,z)$ and $n_{\rm
  L}(L,z)$ for larger clusters most of which would have formed only
recently ($z_{\rm f} \sim z \ll 1$), it is likely to affect the
results for less massive clusters.

This line of consideration motivates us to specify explicitly the
formation epoch of virialized structures and their subsequent
evolution in making statistical predictions for comparison with
observations. A key quantity for this purpose is a distribution
function of halo formation epochs proposed by Lacey \& Cole (1993,
hereafter LC) and Kitayama \& Suto (1996, hereafter KS) in a similar
but slightly different manner (see also Blain \& Longair 1993; Sasaki
1994). In this paper, we combine these distribution functions with a
simple model of cluster gas properties to make quantitative
predictions on the temperature and luminosity functions of clusters of
galaxies in cold dark matter (CDM) universes with/without a
cosmological constant $\lambda_0$.

The plan of this paper is as follows. Section 2 outlines the formalism
we use in computing the temperature and luminosity functions. Section
3 describes a model of X-ray clusters, and our main results are
presented in Section 4. Finally, Section 5 summarizes our conclusions.

\section{\normalsize\bf Formulation}
\subsection{\normalsize\it Temperature and luminosity functions}
\setcounter{equation}{0}

The temperature function $n_{\rm T}(T,z)$ of X-ray clusters is defined
as the differential comoving number density of clusters of temperature
$T$ at a given redshift $z$ (the luminosity and mass functions,
$n_{\rm L}(L,z)$ and $n_{\rm M}(M,z)$, are defined in a similar
manner). On the basis of the spherical collapse model (see Section 3.1
and Appendix A), we assume that the temperature $T$ and the luminosity
$L$ of X-ray clusters are determined by the mass $M$, the redshift of
formation $z_{\rm f}$ and the redshift of observation $z$;
i.e. $T=T(M,z_{\rm f},z)$ and $L=L(M,z_{\rm f},z)$.  Then a proper
theoretical prediction for $n_{\rm T}(T,z)$ and $n_{\rm L}(L,z)$
requires a quantity $F(M,z_{\rm f}; z)$, the number density of objects
of mass $M$ which {\it formed} at $z_{\rm f}$ and are {\it observed}
at $z$, rather than simply a mass function $n_{\rm M}(M,z)$. This is
because $T$ and $L$ depend not only on $M$ and $z$ but also on $z_{\rm
  f}$. Once $F(M,z_{\rm f};z)$ is given, the temperature and
luminosity functions are written respectively as
\begin{equation} 
  n_{\rm T}(T,z) = \left. \int_{z}^{\infty} d z_{\rm f} F(M,z_{\rm
      f};z)\frac{d M}{d T}\right|_{M=M(T,z_{\rm f},z)} ,
\label{tempfnf}
\end{equation}
and
\begin{equation} 
  n_{\rm L}(L,z)  = \left. \int_{z}^{\infty} d z_{\rm f} F(M,z_{\rm
      f}; z) \frac{d M}{d L} \right|_{M=M(L,z_{\rm f},z)} . 
\label{lumfnf}
\end{equation}

By contrast, conventional approaches simply translate the PS mass
function as (e.g. Evrard \& Henry 1991; Hanami 1993)
\begin{equation} 
n_{\rm T}(T,z)  = 
n_{\rm PS}(M,z) \left.\frac{d M}{d T}\right|_{M=M(T,z)}, 
\qquad \hspace{-5mm}
n_{\rm L}(L,z)  = 
n_{\rm PS}(M,z) \left.\frac{d M}{d L}\right|_{M=M(L,z)}, 
\label{tempfnps}
\end{equation}
which correspond to assuming that each cluster forms when it is
observed ($z_{\rm f} = z$). In the above, the PS mass function $n_{\rm
  PS}(M,z)$ is given by
\begin{equation}
\label{eq:psmass}
n_{\rm PS}(M,z) = \sqrt{2 \over \pi} {\rho_\0 \over M} {\delta_{\rm
    c}(z) \over \sigma^2(M)} \abs{d \sigma(M) \over d M} \exp \left[ -
  { \delta_{\rm c}^2(z) \over 2 \sigma^2(M)} \right],
\end{equation}
where $\rho_\0$ ($\simeq 2.78 \times 10^{11} \Omega_0 h^2 \msun$
Mpc$^{-3}$) is the mean comoving density of the universe,
$\sigma^2(M)$ is the mass variance of linear density fluctuations at
the present epoch, and $\delta_{\rm c}(z)$ is the critical linear
overdensity evaluated at present for a spherical perturbation to
collapse at $z$. In what follows, $\sigma^2(M)$ and $\delta_{\rm
  c}(z)$ are computed according to the formulae presented in
Appendices A and B.

Since the conventional PS approach (eq.~[\ref{tempfnps}]) identifies
the epoch of formation $z_{\rm f}$ with that of observation $z$, the
temperature and luminosity evolution of individual clusters after
their formation are not properly taken into account. Our method
(eqs~[\ref{tempfnf}] and [\ref{lumfnf}]), on the other hand, can in
principle include the evolution more naturally. For this purpose, one
needs an appropriate expression for $F(M,z_{\rm f};z)$, as well as an
evolution model which specifies $T=T(M,z_{\rm f},z)$ and $L=L(M,z_{\rm
  f},z)$.  We will discuss these points in \S 2.2 and \S 3.1 below.

\subsection{\normalsize\it Distribution function of halo formation epochs}

\begin{figure}
\begin{center}
   \leavevmode\psfig{figure=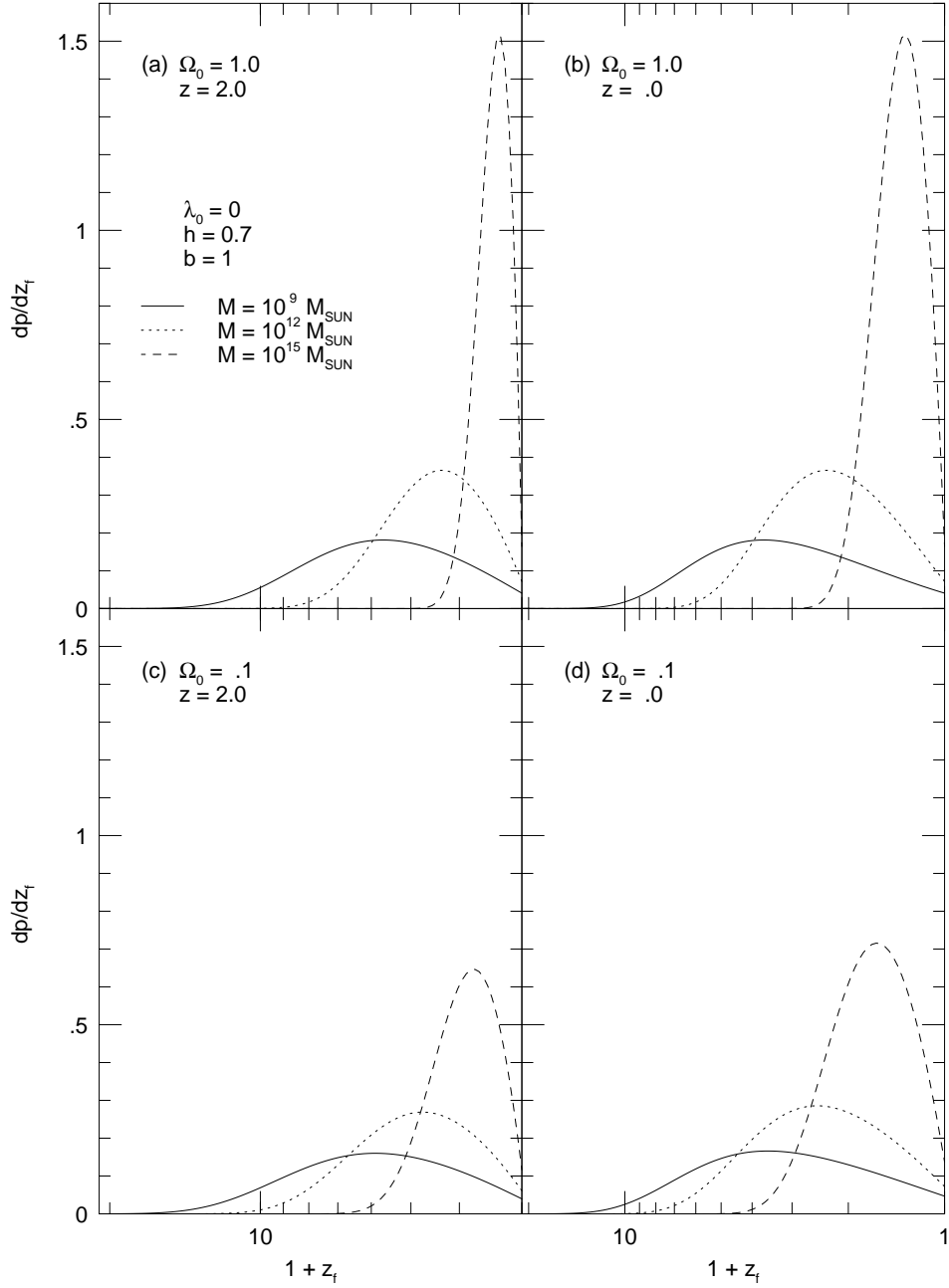,height=17cm}
\end{center}
\caption{The formation epoch distribution function
  $\partial p/\partial z_{\rm f}$ by LC in CDM models; (a)
  $\Omega_0=1$, $z=2$, (b) $\Omega_0=1$, $z=0$, (c) $\Omega_0=0.1$,
  $z=2$, (d) $\Omega_0=0.1$, $z=0$. In all panels, $\lambda_0=0$,
  $h=0.7$, and $b=1$.  Lines indicate $M=10^9 \msun$ (solid), $10^{12}
  \msun$ (dotted), and $10^{15} \msun$ (dashed).}
\label{fig:dpdz}
\end{figure}

In the hierarchical clustering scenario, each halo increases its mass
via major mergers and steady accretion, and thus the formation epoch
$z_{\rm f}$ of a halo is not always well-defined.  LC proposed a
differential distribution function of halo formation epochs, $\partial
p/\partial z_{\rm f}$, defined via the probability that a halo of mass
$M$ which exists at $z$ has a mass greater than $M/2$ for the first
time at $z_{\rm f}$ (see LC \S 2.5.2):
\begin{eqnarray}
\label{eq:dpdzf}
{\partial p \over \partial z_{\rm f}}(M, z_{\rm f}, z) &=&
{\partial p \over \partial\tilde \omega_{\rm f}}(M, \tilde\omega_{\rm
  f}) {\partial \tilde \omega_{\rm f} \over \partial z_{\rm f}},\\ 
\label{eq:dpdwf}
{\partial p \over \partial \tilde \omega_{\rm f}}(M, \tilde\omega_{\rm
  f}) &=& \hspace{-3mm}{1 \over \sqrt{2\pi}} \int_0^1 d\tilde S {M
  \over M'(\tilde S)} {1 \over \tilde S^{3/2}}\left(1-{\tilde
    \omega_{\rm f}^2 \over \tilde S} \right) {\rm exp}\left(-{\tilde
    \omega_{\rm f}^2 \over 2\tilde S} \right),
\end{eqnarray}
where
\begin{eqnarray}
  \tilde \omega_{\rm f}(M, z_{\rm f},z) &\equiv& \frac{\delta_{\rm
      c}(z_{\rm f})-\delta_{\rm
      c}(z)}{\sqrt{\sigma^2(M/2)-\sigma^2(M)}} , \\ 
\label{eq:tildes}
  \tilde S (M',M)&\equiv& {\sigma^2(M')-\sigma^2(M) \over
    \sigma^2(M/2)-\sigma^2(M)} ,
\end{eqnarray}
and $M'(\tilde S)$ is computed by solving equation (\ref{eq:tildes})
for $M'$. Figure \ref{fig:dpdz} shows $\partial p/\partial z_{\rm f}$
as a function of $z_{\rm f}$ in CDM models with $\lambda_0=0$, $h=0.7$
and $b=1$, where $b$ is the bias parameter defined by $b \equiv
1/\sigma(8h^{-1}\mbox{Mpc})$.

Note that, in the above definition of the halo formation epochs, the
mass of a halo at $z_{\rm f}$ and $z$ is different at most by a factor
of 2. Supposedly this factor will be close to 2, since the increase of
halo masses is expected to be dominated by steady accretion of small
objects (although the mass increase is not necessarily continuous
because of major mergers, the fraction of such events should be
relatively small). Thus the LC proposal implies that the quantity
$F(M,z_{\rm f};z)$ in equations (\ref{tempfnf}) and (\ref{lumfnf}) can
be written as
\begin{equation}
  F_{\rm LC}(M,z_{\rm f};z) dM dz_{\rm f}\equiv 2 {\partial p \over
    \partial z_{\rm f}}(2M, z_{\rm f},z) n_{\PS}(2M,z)dM dz_{\rm f} ,
\label{eq:LCF}
\end{equation}
where $M$ is the halo mass at $z_{\rm f}$, which is assumed to have
doubled by $z$, and $n_{\rm PS}(M,z)$ is the PS mass function given by
equation (\ref{eq:psmass}).

As discussed by LC, however, the probability $\partial p/\partial
\tilde \omega_{\rm f}$ given by equation (\ref{eq:dpdwf}) becomes
negative for small $\tilde \omega_{\rm f}$ in the case of power-law
matter fluctuation spectra $P(k)\propto k^n$ with index $n>0$,
implying that the above definition of the halo formation distribution
function is not completely self-consistent.  Since the effective power
index of the observed fluctuation spectrum is negative below the
cluster scale (e.g. Peacock \& Dodds 1994), this would not be a
serious problem for most astrophysically interesting
objects. Nevertheless this clearly exhibits the importance of
exploring other prescriptions for computing the distribution of halo
formation epochs.

One such alternative is given by KS. They derived expressions for the
rates of formation and destruction of bound virialized objects based
upon the conditional probability argument developed by Bond et
al. (1991), Bower (1991), and LC. Their formalism yields the number
density of haloes of mass $M \sim M+dM$ that form via major mergers at
$z_{\rm f}\sim z_{\rm f}+ dz_{\rm f}$ and remain in the range $M \sim
2 M$ at a later time $z$ (see KS \S 2.4):
\begin{eqnarray}
  F_{\rm KS}(M,z_{\rm f};z)dMdz_{\rm f}&=&
  \frac{1}{\sqrt{2\pi}}\frac{1}{\sqrt{\sigma^2(M/2)-\sigma^2(M)}}
  \left\{2-\sbkt{\frac{\delta_{\rm c}(z_{\rm f})-2\delta_{\rm
          c}(z)}{\delta_{\rm c}(z)}} \right.\nonumber \\ 
  &&\hspace{-8mm} \times \exp\lbkt{\frac{2\delta_{\rm
        c}(z)\{\delta_{\rm c}(z_{\rm f})-\delta_{\rm
        c}(z)\}}{\sigma^2(M)}} {\rm erfc}[X(M,z_{\rm f},z)] \nonumber
  \\ && \hspace{-8mm} - {\rm erfc}[Y(M,z_{\rm f},z)]
  \left.\begin{array}{c}\\\\\end{array}\hspace{-4mm}\right\}
  \lbkt{\frac{d\delta_{\rm c}(z_{\rm f})}{dz_{\rm f}}} n_{\rm
    PS}(M,z_{\rm f})dM dz_{\rm f} ,
\label{eq:KSF}
\end{eqnarray}
where erfc($u$) is the complimentary error function defined by    
\begin{equation}
    {\rm erfc}(u)\equiv 
       \frac{2}{\sqrt{\pi}}\int_{u}^{\infty}\exp(-t^{2})dt, 
\end{equation}
and $X(M,z_{\rm f},z)$ and $Y(M,z_{\rm f},z)$ are respectively
\begin{equation}
  X(M,z_{\rm f},z)\equiv\frac{\sigma^{2}(2M) [\delta_{\rm c}(z_{\rm
      f})-2\delta_{\rm c}(z)]+\sigma^2(M)\delta_{\rm c}(z)}
  {\sqrt{2\sigma^2(M)\sigma^2(2M)[\sigma^2(M)-\sigma^{2}(2M)]}},
\end{equation}
\begin{equation}
  Y(M,z_{\rm f},z)\equiv \frac{\sigma^2(M)\delta_{\rm
      c}(z)-\sigma^{2}(2M)\delta_{\rm c}(z_{\rm f})}
  {\sqrt{2\sigma^2(M)\sigma^2(2M)[\sigma^2(M)-\sigma^{2}(2M)]}}.
\end{equation}
As mentioned by KS, the above definition (eq.~[\ref{eq:KSF}]) for the
formation epoch distribution may lead to a systematic overestimation
of the number of haloes that form by mergers at $z_{\rm f}$ (see \S 4
of KS).

\begin{figure}
\begin{center}
   \leavevmode\psfig{figure=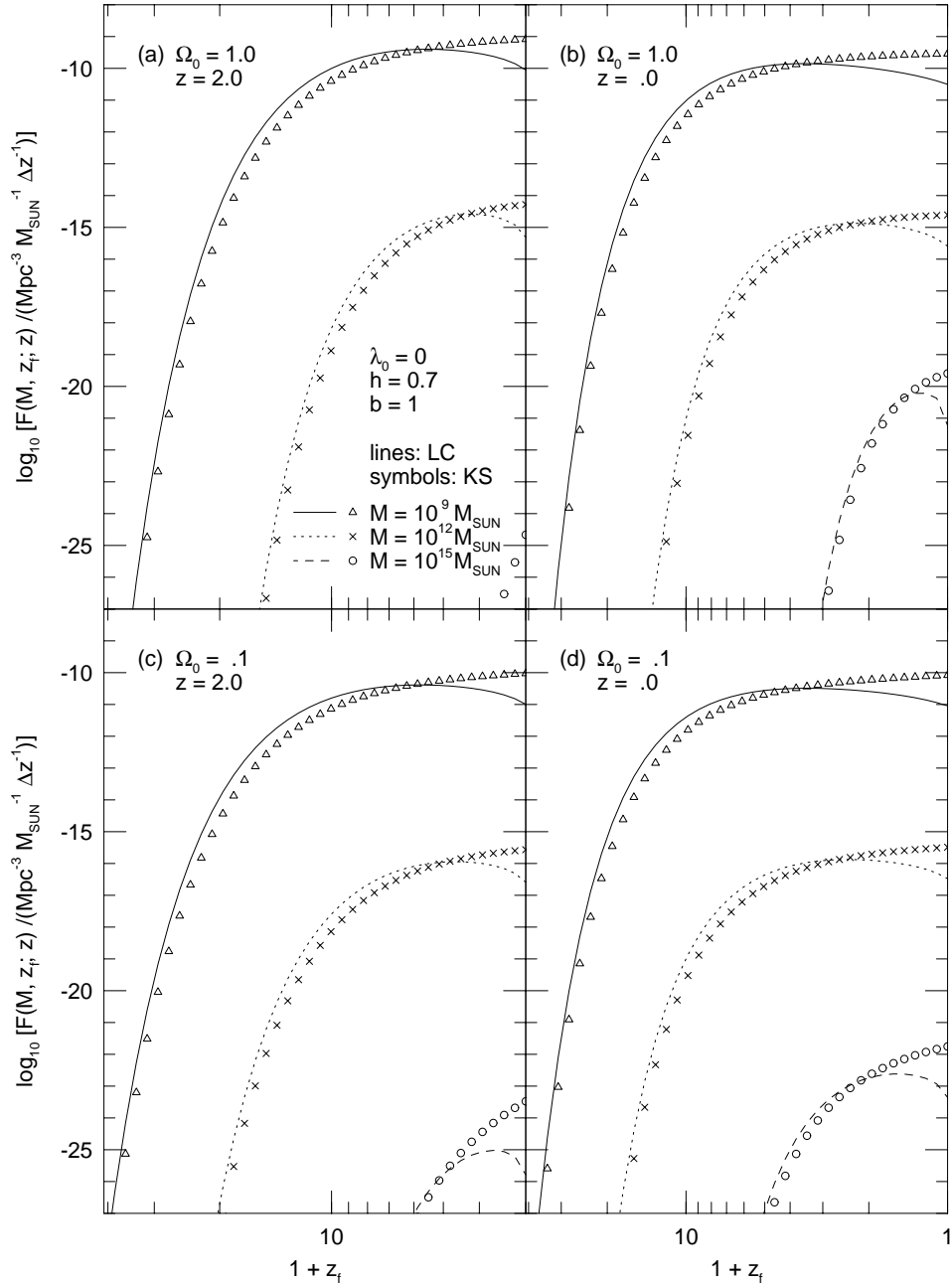,height=17cm}
\end{center}
\caption{The distribution function of halo formation epochs by 
  LC (eq.~[\protect\ref{eq:LCF}\protect]) and KS
  (eq.~[\protect\ref{eq:KSF}\protect]) in CDM models; (a)
  $\Omega_0=1$, $z=2$, (b) $\Omega_0=1$, $z=0$, (c) $\Omega_0=0.1$,
  $z=2$, (d) $\Omega_0=0.1$, $z=0$. In all panels, $\lambda_0=0$,
  $h=0.7$, $b=1$.  Lines and symbols indicate the LC and KS formulae
  respectively for $M=10^{9}\msun$ (solid line, triangles),
  $10^{12}\msun$ (dotted line, crosses), and $10^{15}\msun$
  (dashed line, circles).}
\label{fig:kslc}
\end{figure}

Strictly speaking, therefore, neither $F_{\rm LC}$ nor $F_{\rm KS}$ is
a completely satisfactory expression for $F(M,z_{\rm f};z)$.  In
practice, however, $F_{\rm LC}$ and $F_{\rm KS}$ are in good agreement
except at $z_{\rm f} \sim z$ (Fig.~\ref{fig:kslc}); the small
discrepancy is ascribed to different criteria of ``formation'' in the
two approaches, because the formation epochs are not uniquely defined
for haloes which increase their mass quiescently via accretion.
Furthermore, as will be shown in \S 4, such a difference between the LC
and KS models does not affect the predictions for $n_{\rm T}(T,z)$ and
$n_{\rm L}(L,z)$ (see Figs~\ref{fig:omgt} and \ref{fig:omgl}). Thus we
believe that they provide reasonable approximations to $F(M,z_{\rm
  f};z)$, at least for our present purpose.

\section{\normalsize\bf Descriptions of X-ray clusters}
\setcounter{equation}{0}

\subsection{\normalsize\it A simple model}
\label{secmodel}

Given the distribution function of halo formation epochs $F(M,z_{\rm
  f};z)$, it remains to specify $T=T(M,z_{\rm f},z)$ and $L=L(M,z_{\rm
  f},z)$.  We basically follow our previous model (\S 3.1.2 of KS),
but slightly modify the hypothesis on the temperature and luminosity
evolution.

We make the approximation that X-ray clusters consist of
dissipationless dark matter and baryonic gas. The dark matter
component dominates the total gravitating mass and is supposed to be
in virial equilibrium. Once the cosmological parameters such as
$\Omega_0$ and $\lambda_0$ are specified, the virial radius $r_{\rm
  vir}$ and the virial temperature $T_{\rm vir}$ are determined by the
mass $M$ and the formation epoch $z_{\rm f}$ using the spherical
collapse model (e.g. Peebles 1980; Appendix A) We compute these
quantities from the mean density $\rho_{\rm vir}$ given in Appendix A
by 
\begin{eqnarray} 
  r_{\rm vir}(M,z_{\rm f}) &=& \sbkt{3M \over 4\pi\rho_{\rm
      vir}}^{1/3}, \\ T_{\rm vir}(M,z_{\rm f}) &=& {G M \mu m_{\rm p}
    \over 3k_{\rm B} r_{\rm vir}},
\end{eqnarray}
where $G$ is the gravitational constant, $k_{\rm B}$ is the Boltzmann
constant, $m_{\rm p}$ is the proton mass, and $\mu$ is the mean
molecular weight. Hereafter we assume that the intracluster gas is
fully ionized with primordial abundances of helium and hydrogen, and
thus set $\mu = 0.59$.

For a cluster temperature of $\ngt 3 $keV, the contribution from the
line emission on the total X-ray luminosity can be neglected.  Since
the shape and amplitude of the temperature and luminosity functions
are derived observationally for relatively high temperature clusters
(see the error box in Figs~\ref{fig:omgt} $\sim$ \ref{fig:z} below;
Henry \& Arnaud 1991), we compute the X-ray luminosity only from
thermal bremsstrahlung emission. Incidentally, the X-ray line emission
from less massive clusters may be important in considering the origin
of the soft X-ray background (Cen et al. 1995; Suto et al. 1996), and
we will discuss the effect of line emissions elsewhere (Sasaki et
al. 1996).

We assume that the cluster gas is isothermal with temperature given by
\begin{eqnarray} 
  T(M,z_{\rm f},z) = \kappa(z_{\rm f},z) T_{\rm vir}(M,z_{\rm f}),
\label{gastemp}
\end{eqnarray} 
where $\kappa(z_{\rm f},z)$ is introduced so that temperature
evolution due to quiescent accretion of matter after $z_{\rm f}$ can
be taken into account. Note that this treatment is possible because
our method explicitly distinguishes $z_{\it f}$ from $z$.  As there is
not yet a good theoretical model to describe the temperature
evolution, we simply suppose that $\kappa(z_{\rm f},z)$ takes a
power-law form:
\begin{equation} 
\kappa(z_{\rm f},z)=\sbkt{\frac{1+z_{\rm f}}{1+z}}^s .
\label{kappa}
\end{equation}
Hydrodynamical simulations indicate that the temperature of clusters,
once formed, does not change drastically after virialization, and are
roughly consistent with $0 \nlt s \nlt 1$ (e.g. Evrard 1990; Navarro,
Frenk \& White 1995).

We adopt the following spherically symmetric distribution for
intracluster gas density (e.g. Jones \& Forman 1984)
\begin{equation}
\rho_{\rm gas}(r)=\rho_{\rm gas}^0 
                      \lbkt{1+\sbkt{\frac{r}{r_{\rm c}}}^2}^{-1},  
\label{densprof}
\end{equation}
where $\rho_{\rm gas}^0$ is the central gas density, and $r_{\rm c}$
is the core radius. Since this is basically an empirical fitting
formula, it is difficult to predict $r_{\rm c}$. Thus we adopt a
simple self-similar model in which $r_{\rm c}$ is proportional to
$r_{\rm vir}(M,z_{\rm f})$:
\begin{eqnarray} 
r_{\rm c} &=& 0.15 h^{-1}\mbox{Mpc } \times \frac{r_{\rm vir}(M,z_{\rm
f})}{r_{\rm vir}(10^{15}\msun, z_{\rm f}=0)}  
\end{eqnarray}
where the normalization is chosen to match the X-ray observations
(Abramopoulos \& Ku 1983; Jones \& Forman 1984).  Now $\rho_{\rm
  gas}^0$ is fixed by the relation:
\begin{equation}
  \int_0^{r_{\rm vir}} \rho_{\rm gas}(r)\, 4\pi r^2 d r = M
  \sbkt{\frac{\Omega_{\rm B}}{\Omega_0}} ,
\end{equation}
where $\Omega_{\rm B}$ is the baryonic density parameter of the
universe chosen to be consistent with primordial nucleosynthesis:
$\Omega_{\rm B} =0.0125 h^{-2}$ (Walker et al. 1991).  We assume no
intrinsic evolution for the gas density profile, which is also
suggested by numerical simulations (e.g. Navarro, Frenk \& White
1995).

Combining all the above assumptions, the temperature $T$ and the
bolometric luminosity $L_{\rm bol}$ of X-ray clusters in our model are
roughly related to $M$, $z_{\rm f}$ and $z$ as follows:
\begin{equation} 
T(M,z_{\rm f},z) 
\propto M^{2/3} (1+z_{\rm f})^{\xi}\sbkt{\frac{1+z_{\rm f}}{1+z}}^s 
\end{equation} 
\begin{equation} 
  L_{\rm bol}(M,z_{\rm f},z) 
\propto M^{4/3} (1+z_{\rm f})^{7\xi/2}\sbkt{\frac{1+z_{\rm f}}{1+z}}^{s/2},
\end{equation} 
where $\xi$ is the effective index which varies weakly with $z_{\rm
  f}$, $\Omega_0$ and $\lambda_0$; e.g. $\xi =1$ independent of
$z_{\rm f}$ in the case of $(\Omega_0,\lambda_0)=(1,0)$, while $\xi
\simeq 0.5$ at $z_{\rm f}\sim 0 $ and $\xi \simeq 1$ at $z_{\rm f}
\ngt 2$ for $(\Omega_0,\lambda_0)=(0.2,0.8)$.

\subsection{\normalsize\it Comparison with observed correlations}

\begin{figure}
\begin{center}
   \leavevmode\psfig{figure=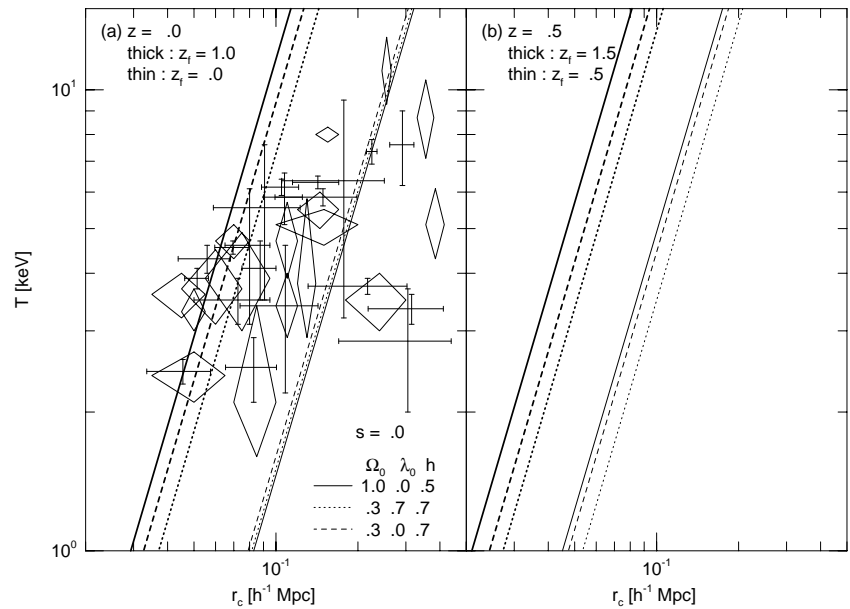,height=9cm,angle=90}
\end{center}
\caption{The $T - r_{\rm c}$ relation for the clusters in 
  CDM models at (a) $z=0$ and (b) $z=0.5$. Three representative
  models, Eh5, L3h7, and O3h7 are plotted in solid, dotted, and dashed
  lines, respectively. Thick and thin lines correspond to the results
  for $z_{\rm f}=z+1$ and $z_{\rm f}=z$, respectively. The crosses
  show the {\it Einstein} observations (Jones \& Forman 1984; David et
  al. 1993), and the diamonds indicate {\it EXOSAT} data (Edge \&
  Stewart 1991a,b).}
\label{fig:trc}
\end{figure}

\begin{figure}
\begin{center}
   \leavevmode\psfig{figure=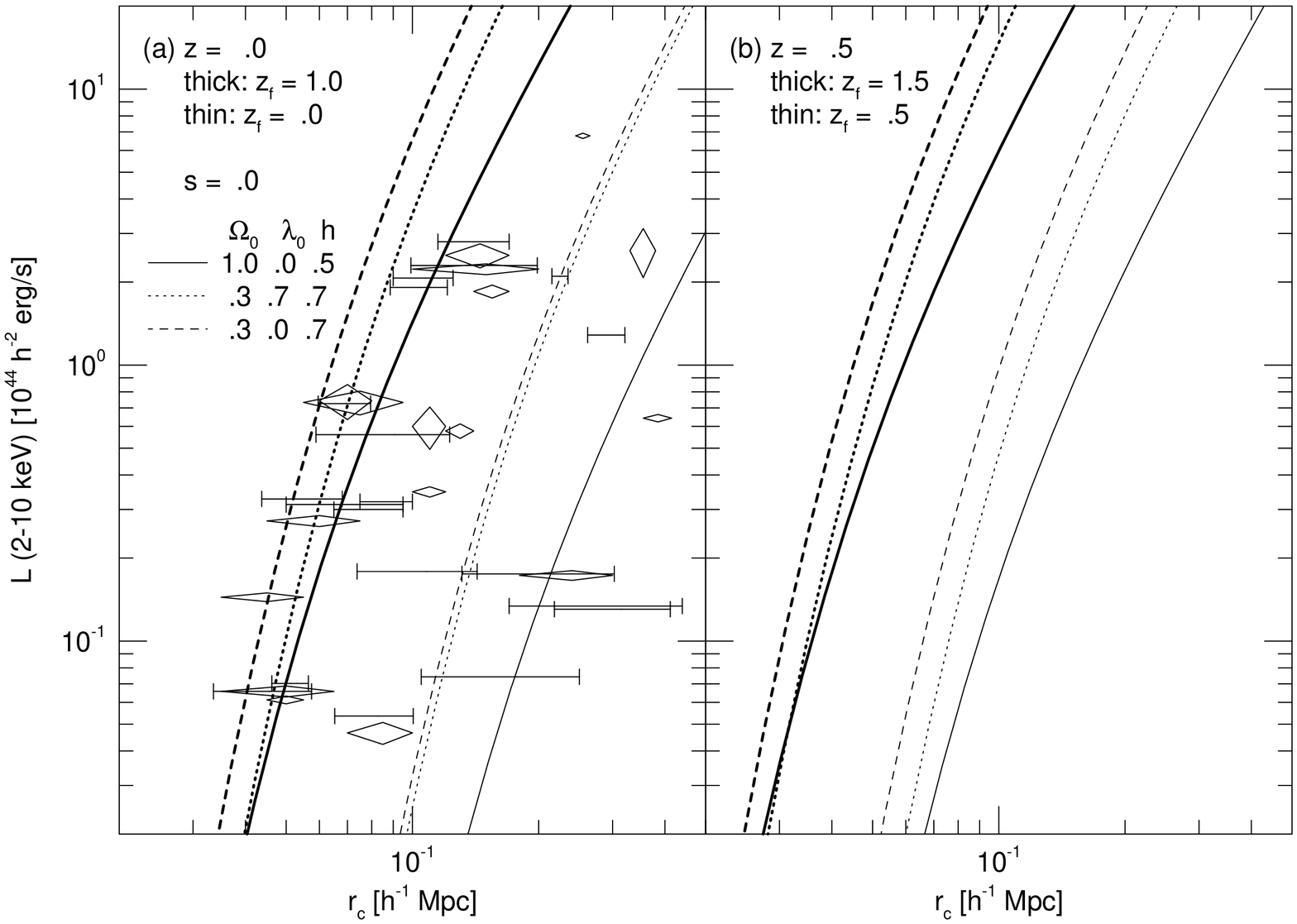,height=8cm,angle=90}
\end{center}
\caption{Same as Figure \protect\ref{fig:trc}\protect~ 
  except for the $L (2-10 {\rm keV}) - r_{\rm c}$ relation.}
\label{fig:lmrc}
\end{figure}

\begin{figure}
\begin{center}
  \leavevmode\psfig{figure=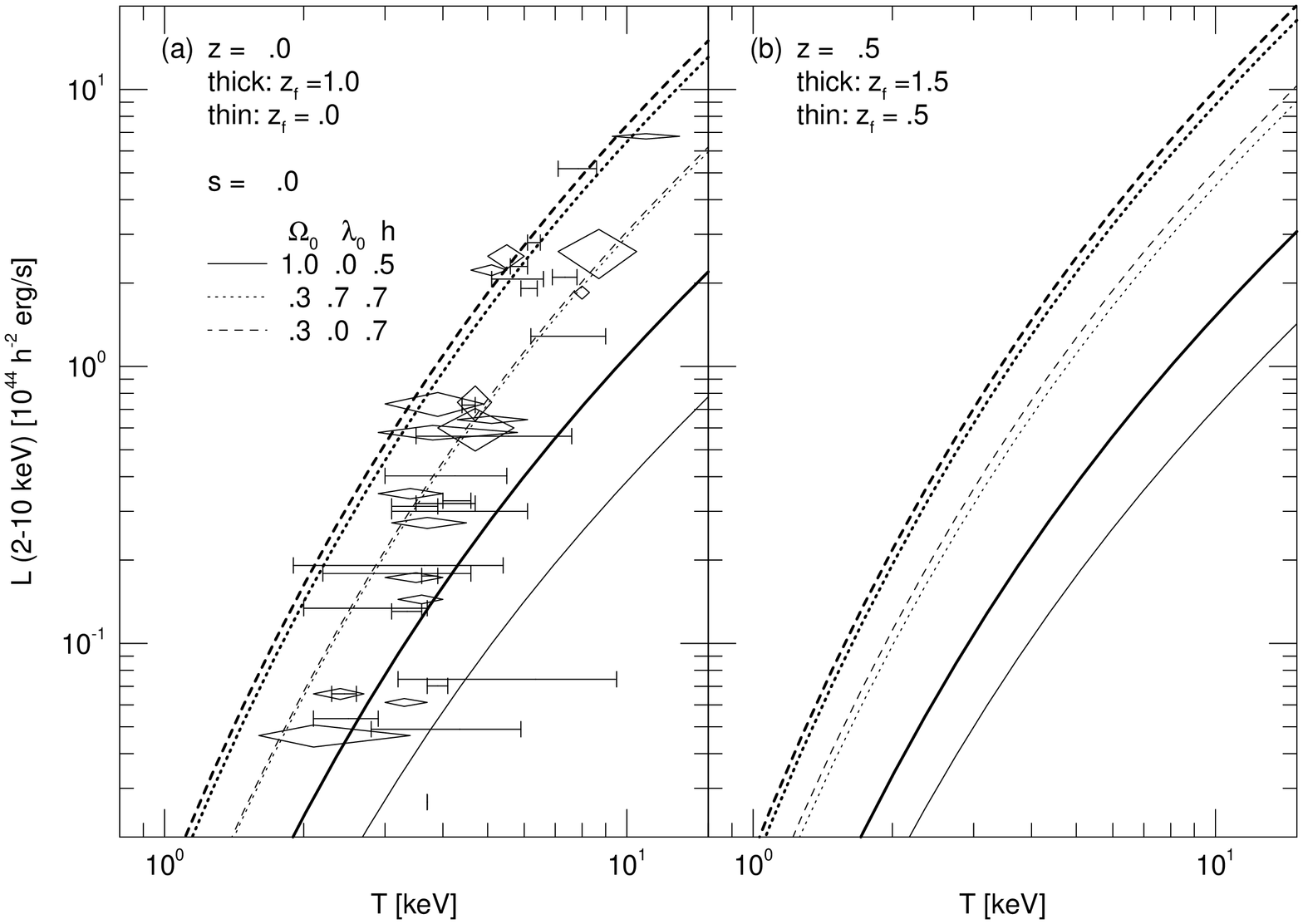,height=8cm,angle=90}
\end{center}
\caption{Same as Figure  \protect\ref{fig:trc}\protect~ 
  except for the $L (2-10 {\rm keV}) - T$ relation.}
\label{fig:lmtp}
\end{figure}

The model presented in the previous subsection is admittedly crude, so
we need to examine the extent to which it is consistent with the
available observations before presenting the results for $n_{\rm
  T}(T,z)$ and $n_{\rm L}(L,z)$.  Specifically, we plot the
correlations among $T$, $r_{\rm c}$ and $L(2-10 {\rm keV})$ in Figures
\ref{fig:trc}, \ref{fig:lmrc} and \ref{fig:lmtp}. In these figures we
select three representative CDM models, Eh5, L3h7, and O3h7, which
have respectively ($\Omega_0$, $\lambda_0$, $h$) = (1.0, 0, 0.5),
(0.3, 0.7, 0.7), and (0.3, 0, 0.7). In fact, model L3h7 is shown to be
reasonably consistent with the observed temperature and luminosity
functions at $z=0$ in the next section.

Basically, our models reproduce fairly well the observed $T - r_{\rm
  c}$ and $L(2-10 {\rm keV}) - r_{\rm c}$ correlations, considering
that most clusters present at $z=0$ would have formed at $z_{\rm
  f}\nlt 1$.  The slopes of our models in Figures \ref{fig:trc} and
\ref{fig:lmrc} would become less steep if the distribution of $z_{\rm
  f}$ is taken into account, because massive clusters (large $T$ and
$L$) are likely to have formed more recently than less massive
ones. As regards the $L(2-10 {\rm keV}) - T$ correlation on the other
hand, only low $\Omega_0$ models show reasonable agreement with the
observed data at large $L$ and $T$. The predicted curves are less
steep than the observed one. In fact, several authors have pointed out
that simple adiabatic models fail to account for the observed $L - T$
relation (e.g. Navarro, Frenk \& White 1995). It has also been
suggested that the discrepancy may possibly be reconciled if the gas
has attained large entropy at high redshift (Kaiser 1991; Evrard 1990;
Evrard \& Henry 1991).

Apparently this is an important area of research, especially if one
takes seriously the above discrepancy between the observed $L - T$
relation and the theoretical prediction of a simple adiabatic model.
Nevertheless it is fairly independent of our current formalism; once
any physical model of such pre-heating specifies $T=T(M,z_{\rm f},z)$
and $L=L(M,z_{\rm f},z)$, then we can compute the resulting $n_{\rm
  T}(T,z)$ and $n_{\rm L}(L,z)$ by a straightforward application of
the method described here.  Thus we do not discuss this problem
further and hope to come back to it elsewhere.

\section{\normalsize\bf Predictions for temperature and luminosity functions}
\setcounter{equation}{0}

\begin{figure}
\begin{center}
   \leavevmode\psfig{figure=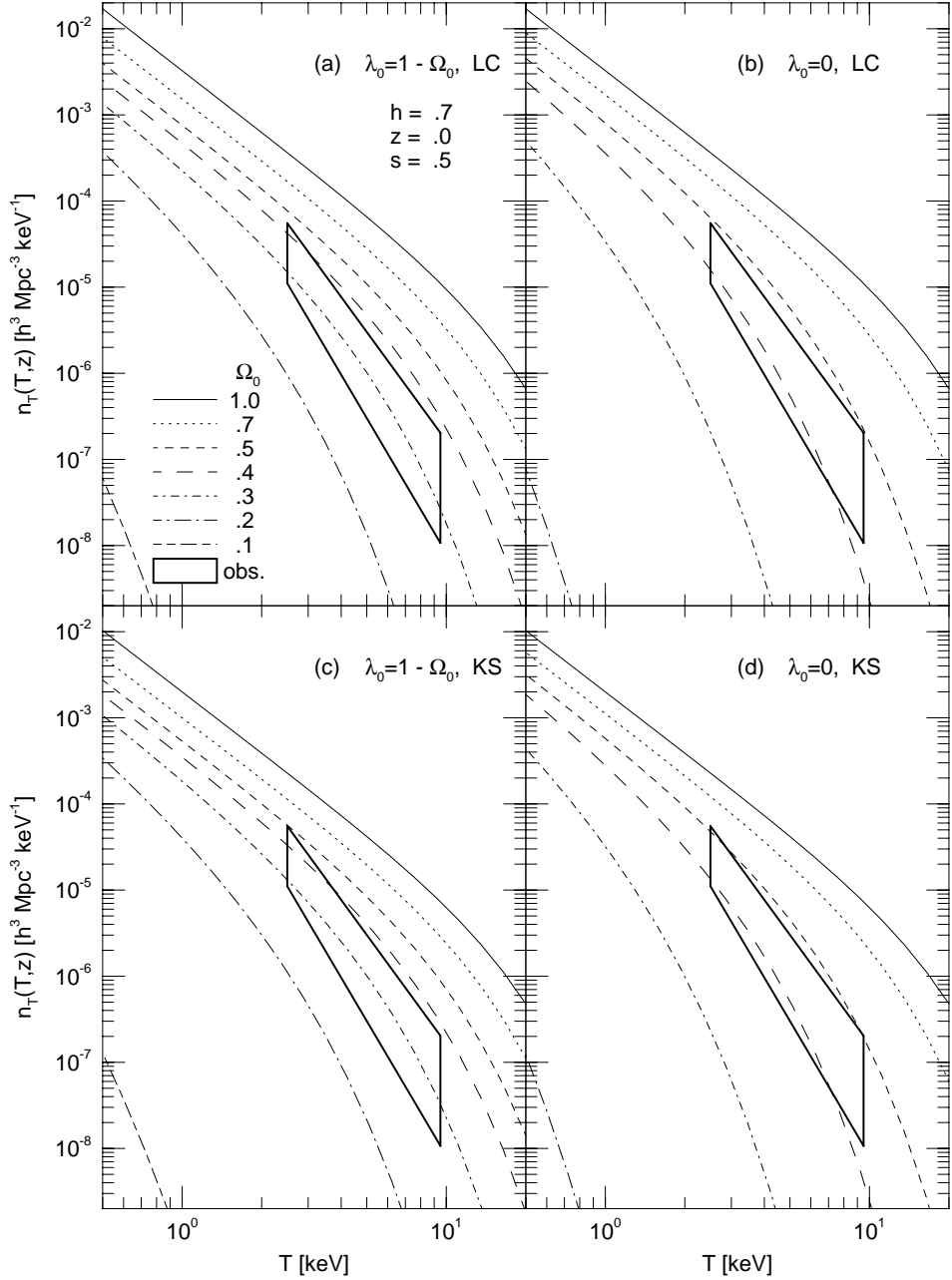,height=17cm}
\end{center}
\caption{Predictions of the temperature function of clusters at $z=0$ 
  in CDM universes with $h=0.7$; (a) LC model with
  $\lambda_0=1-\Omega_0$, (b) LC model with $\lambda_0=0$, (c) KS
  model with $\lambda_0=1-\Omega_0$, and (d) KS model with
  $\lambda_0=0$. The fluctuation amplitudes are normalized according
  to the {\it COBE} 2 year data (Gorski et al. 1995; Sugiyama
  1995). The thick solid box indicates the observational error box
  (Henry \& Arnaud 1991).}
\label{fig:omgt}
\end{figure}

\begin{figure}
\begin{center}
   \leavevmode\psfig{figure=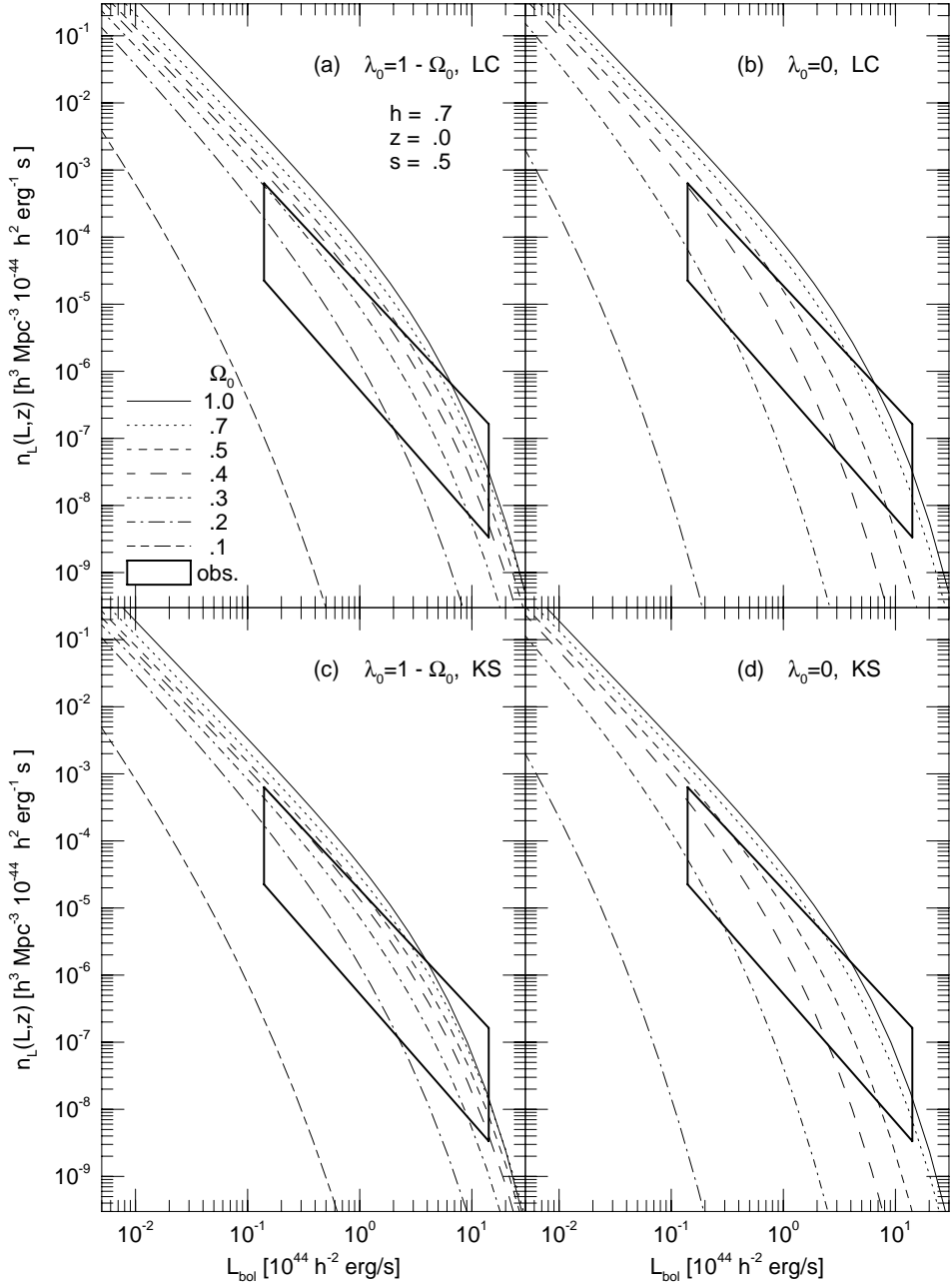,height=17cm}
\end{center}
\caption{Same as Figure \protect\ref{fig:omgt}\protect~  except 
  for the bolometric luminosity function.}
\label{fig:omgl}
\end{figure}

The temperature and luminosity functions of clusters predicted in CDM
universes at $z=0$ are shown in Figures \ref{fig:omgt} and
\ref{fig:omgl}, respectively. All the models adopt $h=0.7$ based on
the recent observations of Cepheids by the Hubble Space Telescope
(Tanvir et al. 1995).  The fluctuation amplitudes are normalized
according to the {\it COBE} 2 year data (Gorski et al. 1995; Sugiyama
1995). Panels (a) and (b) in these figures are based on the LC model
(eq.~[\ref{eq:LCF}]) while panels (c) and (d) adopt the KS model
(eq.~[\ref{eq:KSF}]).  The comparison implies that the results are
quite insensitive to the choice of models, because the small
difference between the two models shown in Figure \ref{fig:kslc} has
been canceled out in the integration over $z_{\rm f}$
(eqs~[\ref{tempfnf}] [\ref{lumfnf}]). Thus in what follows we will
adopt the LC model just for definiteness.  Clearly the predictions of
the temperature and luminosity functions vary largely with the shape
and the amplitude of the fluctuation spectrum. Since both of these are
fixed by $\Omega_0$ and $\lambda_0$ assuming the CDM model and the
{\it COBE} normalization, comparison of the theoretical predictions
with the observed temperature and luminosity functions serves as a
possible discriminator of the values of $\Omega_0$ and $\lambda_0$.

\begin{figure}
\begin{center}
   \leavevmode\psfig{figure=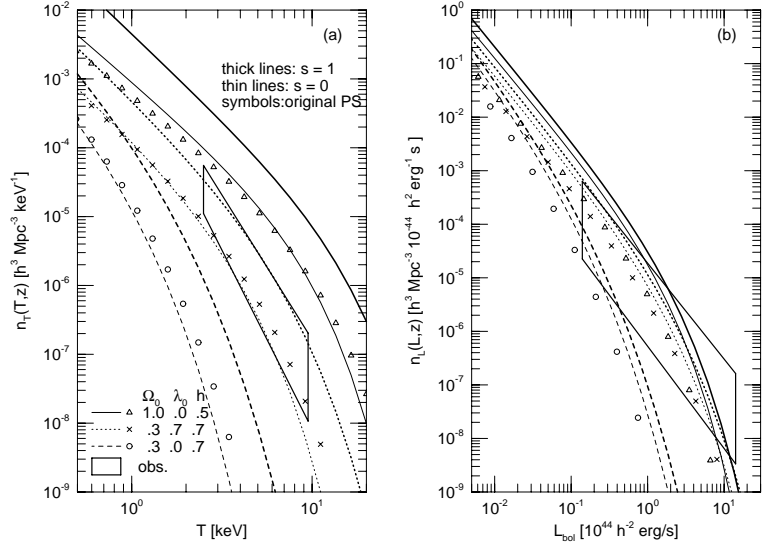,height=8cm,angle=90}
\end{center}
\caption{Effect of different evolutionary assumptions on the 
  predictions of (a) temperature and (b) luminosity functions at $z=0$
  in Eh5, L3h7, and O3h7 CDM models ({\it COBE} normalized). Results
  for the $s=1$ and $s=0$ models are respectively plotted in thick and
  thin lines, while those based on the original PS model are plotted
  in symbols.}
\label{fig:tev}
\end{figure}

We have tested the robustness of the above predictions against our
assumptions about the temperature evolution in Figure
\ref{fig:tev}. The temperature functions depend rather sensitively on
the evolution index $s$. In fact, the temperature evolution assumed in
our model, $T(z)/T(z_{\rm f}) = (1+z_{\rm f})^s/(1+z)^s$ with $s=1$
(thick lines), tends to move the predictions of $n_{\rm T}(T,z=0)$ to
the right roughly by a factor of $(1+z_{\rm f})$ relative to the $s=0$
case (thin lines). Although the formation epoch $z_{\rm f}$ varies
with $M$ and therefore with $T$ of an individual cluster, $\langle
z_{\rm f}\rangle \sim 0.5$ for typical clusters reasonably accounts
for the behavior in Figure \ref{fig:tev}a. On the other hand, the
shift in the luminosity function is rather small (Figure
\ref{fig:tev}b), because our evolution model assumes that the cluster
density is fairly constant after virialization and the bolometric
luminosity evolves as $L_{\rm bol}(z) /L_{\rm bol}(z_{\rm f})=
(1+z_{\rm f})^{s/2}/(1+z)^{s/2}$.

Incidentally, Figure \ref{fig:tev} shows that our $s=0$ model
prediction is very close to the original PS approach
(eq.~[\ref{tempfnps}]), especially for the temperature
function. However, this does not imply that the effect of $z_{\rm f}
\not= z$ is negligible. Our predictions have been shifted relative to
the original PS approach in the following two ways. One is the
horizontal shift due to the different hypotheses for the temperature
and luminosity evolution; the original PS approach implicitly assumes
for instance $T(z)/T(z_{\rm f})=(1+z)/(1+z_{\rm f})$ and
$L(z)/L(z_{\rm f})=(1+z)^{7/2}/(1+z_{\rm f})^{7/2}$ in the
$\Omega_0=1$ universe, while we assume those just mentioned in the
last paragraph. The other is the vertical shift due to the different
mass of haloes at $z_{\rm f}$ and at $z$; our formalism assumes that
the mass of a halo at $z_{\rm f}$ is half of that at $z$, while no
mass change is considered in the original PS approach. The former
shift depends on the evolution index $s$, and the latter one is
sensitive to the fluctuation spectral index. On cluster scales, the
above effects have almost canceled each other for the temperature
function with $s=0$.

In any case, Figures \ref{fig:omgt}, \ref{fig:omgl} and \ref{fig:tev}
indicate that, with the {\it COBE} normalization, a change in the
parameter $s$ in the range $0 \nlt s \nlt 1$ roughly corresponds to
changing $\Omega_0$ by $0.1\sim 0.2$. In the case of $s=0$ and
$h=0.7$, either the $(\Omega_0, \lambda_0) \simeq (0.4, 0.6)$ or the
$(0.5,0)$ CDM model fit best within the observational error box. If
$s=1$ is adopted instead, models with $(\Omega_0,\lambda_0)\simeq
(0.3,0.7)$ or $(0.4,0)$ are prefered. Of course, the conventional
argument of the age problem also points to a lower $\Omega_0$; if
$h=0.7$ as we adopted here, the ages of the universe are 13.5 Gyr,
12.4 Gyr, 10.9 Gyr and 10.5 Gyr for $(\Omega_0, \lambda_0)=(0.3,
0.7)$, $(0.4, 0.6)$, $(0.4,0)$ and $(0.5,0)$ respectively.

\begin{figure}
\begin{center}
   \leavevmode\psfig{figure=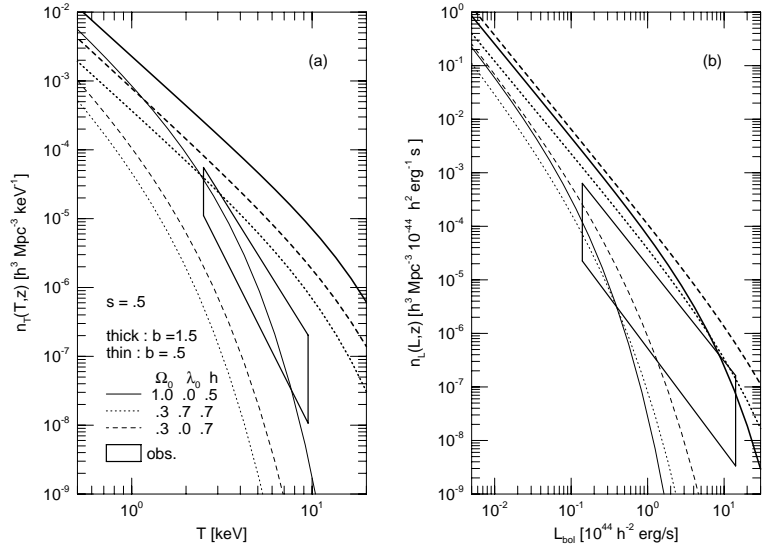,height=8cm,angle=90}
\end{center}
\caption{Effect of fluctuation amplitudes on predictions
  of (a) temperature and (b) luminosity functions at $z=0$ in Eh5,
  L3h7, and O3h7 CDM models ($s=0.5$).  Thick and thin lines indicate
  the results for $b=1.5$ and $b=0.5$, respectively.}
\label{fig:bias}
\end{figure}

Adopting the {\it COBE} normalization in the computation of the
temperature and luminosity functions, we implicitly assume that
luminous clusters trace the underlying mass distribution.  Although
this seems quite reasonable on scales relevant for formation and
clustering of galaxy clusters, it is not easy to rigorously justify
this assumption. So we change the normalization of the fluctuations
and plot the results for $b=1.5$ and $0.5$ in Figure \ref{fig:bias}
(assuming $s=0.5$). For comparison, the {\it COBE} 2 year data imply
that $b=0.73$, $0.87$, and $1.8$ in models Eh5, L3h7 and O3h7
respectively.

With the above uncertainty and the observational error in mind, an
acceptable range of $\Omega_0$ from the observed temperature and
luminosity functions is $0.2 \simlt \Omega_0 \simlt 0.4$ in the case
of spatially flat CDM universe, and $0.3 \simlt \Omega_0 \simlt 0.5$
for open CDM universe. It should be noted that the $\Omega_0 =1$ CDM
model fails to account for the observation by a wide margin and can be
ruled out even from the present comparison only.

\begin{figure}
\begin{center}
   \leavevmode\psfig{figure=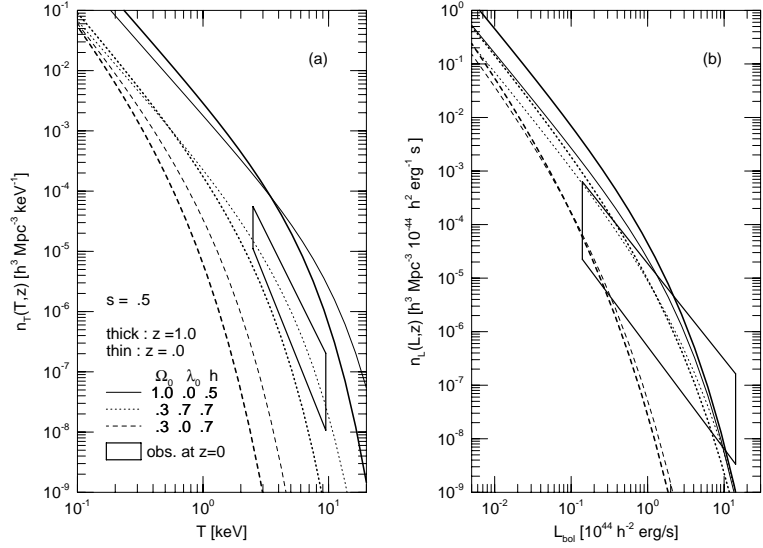,height=8cm,angle=90}
\end{center}
\caption{Evolution of (a) temperature and (b) luminosity
  functions in Eh5, L3h7, and O3h7 CDM models ($s=0.5$). Thick and
  thin lines indicate the results for $z=1$ and $z=0$, respectively.}
\label{fig:z}
\end{figure}

More interesting for the future X-ray observations are the predictions
of evolutionary behaviors in the temperature and luminosity functions.
Figure \ref{fig:z} plots the predictions at $z=0$ and $1$.  Evolution
of the temperature function (Figure \ref{fig:z}a) is in general
negative (i.e. decreasing number density toward the past) for massive
clusters (large $T$), while it becomes positive for smaller ones.  In
model L3h7, for instance, a significant negative evolution occurs at
$T\ngt 2$keV, where observed data are available at $z\sim 0$. On the
other hand, the luminosity function mostly shows very weak evolution
in the range of observational interest (Figure \ref{fig:z}b). This is
still consistent, within the observational error bar, with the results
by Gioia et al. (1990) which indicate weakly negative evolution at
$L(0.3-3.5 \mbox{keV}) \ngt 10^{44} h^{-2} \mbox{erg s}^{-1}$. Our
model can be tested against future observations through the different
evolutionary signature of $n_{\rm T}(T,z)$ (negative evolution) and
$n_{\rm L}(L,z)$ (almost no evolution).

\section{\normalsize\bf Conclusions} 
\setcounter{equation}{0}

We have examined statistical properties of X-ray clusters in CDM
universes in a semi-analytic manner, which is complementary to
numerical simulations. Our method is different from the previous
approaches based upon the conventional PS theory (e.g. Evrard \&
Henry; Hanami 1993), in that we explicitly took account of the epochs
of cluster formation adopting the halo formation epoch distributions
proposed by LC and KS. As a result, our method can include the
subsequent evolution of cluster temperature and luminosity.

Although the LC and KS proposals involve slightly different
definitions of the halo formation epochs, we found that the resultant
temperature and luminosity functions are remarkably similar.
Deviations from the previous PS approach become larger for clusters
that form at higher redshift and for stronger evolution.

The shape and amplitude of the temperature and luminosity functions
vary sensitively with the cosmological density parameter $\Omega_0$,
if the fluctuation amplitude is fixed by the {\it COBE} normalization.
Given the qualitative nature of our simple model, however, we should
not constrain the cosmological parameters so stringently. Rather we
conclude that the low-density $\Omega_0 \sim 0.2 - 0.5$ CDM models
with/without the cosmological constant are roughly consistent with the
observed temperature and luminosity functions. Nevertheless we can
argue against $\Omega_0 =1$ CDM models at least from the present
analysis only.

\vskip1.2cm
\noindent
We thank Takahiro T. Nakamura and Naoshi Sugiyama for useful
discussions on the spherical collapse model and the {\it COBE}
normalization. We are grateful to Ewan D. Stewart for a careful
reading of the manuscript. This research was supported in part by the
Grants-in-Aid by the Ministry of Education, Science and Culture of
Japan (07740183, 07CE2002).

\vfill\eject
\centerline{\bf APPENDICES}

\appendix
\renewcommand{\thesection}{\normalsize\bf\Alph{section}}
\renewcommand{\thesubsection}
{\normalsize\it\Alph{section}.\normalsize\it\arabic{subsection}. }
\renewcommand{\theequation}{\mbox{\rm {\Alph{section}.\arabic{equation}}}}

\section{\normalsize\bf Spherical collapse in $\Omega_0 \leq 1$ universes}
\setcounter{equation}{0} 
\label{sph}

In the present analysis, we need the mean density $\rho_{\rm
  vir}(z_{\rm f})$ and the critical linear overdensity $\delta_{\rm
  c}(z_{\rm f})$ of an object that virializes at $z_{\rm f}$ in a
universe with arbitrary $\Omega_0$ ($\Omega \leq 1$).  For
definiteness, we summarize the results of the spherical collapse model
as well as the fitting formula we used for $\lambda_0=1-\Omega_0$
models (Peebles 1980; Lahav et al. 1991; LC; Nakamura 1996). Note that
$\delta_{\rm c}$ given below is the value computed at virialization,
and $\delta_{\rm c}(z_{\rm f})$ used in our main text is the value
computed at present. They are related via $\delta_{\rm c}(z_{\rm
  f})\equiv \delta_{\rm c}\cdot D(z_{\rm f}=0)/D(z_{\rm f})$, where
$D(z)$ is the linear growth rate.

\begin{enumerate}
\item $\Omega_0=1$, 
\begin{eqnarray}
  \frac{\rho_{\rm vir}(z_{\rm f})}{\bar{\rho}(z_{\rm f})} &=& 18 \pi^2
  \simeq 178,\\ 
 \delta_{\rm c} &=& {3(12\pi)^{2/3} \over 20} \simeq 1.69,
\end{eqnarray}

\item $\Omega_0 <1$, $\lambda_0=0$ ,
\begin{eqnarray}
 && \frac{\rho_{\rm vir}(z_{\rm f})}{\bar{\rho}(z_{\rm f})} = 4 \pi^2
  \frac{(\cosh \eta_{\rm f} - 1)^3}{(\sinh \eta_{\rm f} - \eta_{\rm
      f})^2},\\ &&\hspace{-12mm} \delta_{\rm c} =
  \frac{3}{2}\lbkt{\frac{3\sinh \eta_{\rm f}(\sinh
      \eta_{\rm f} - \eta_{\rm f})}{(\cosh \eta_{\rm f}-1)^2}-2}
  \lbkt{1+\sbkt{\frac{2\pi}{\sinh \eta_{\rm f} - \eta_{\rm
          f}}}^{2/3}},
\end{eqnarray}

\item $\Omega_0 <1$, $\lambda_0=1-\Omega_0$ , 
\begin{eqnarray}
  \frac{\rho_{\rm vir}(z_{\rm f})}{\bar{\rho}(z_{\rm f})} &=&
  \sbkt{\frac{r_{\rm ta}}{r_{\rm vir}}}^3 \frac{2 w_{\rm
      f}}{\chi},\nonumber \\ &\simeq& 18 \pi^2 (1+0.4093 w_{\rm
    f}^{0.9052}), \\ 
  \delta_{\rm c} &=& \frac{3}{5} F\sbkt{{1 \over 3}, 1, {11 \over 6}; -
    w_{\rm f}} \sbkt{\frac{2w_{\rm f}}{\chi}}^{1/3} \sbkt{1+{\chi
      \over 2}}, \nonumber \\ &\simeq& {3(12\pi)^{2/3} \over 20}
  (1+0.123 \log_{10}\Omega_{\rm f}),
\end{eqnarray}
\end{enumerate}

\noindent 
In the above, $\bar{\rho}(z_{\rm f}) \equiv \rho_0 (1+z_{\rm f})^3$,
$\eta_{\rm f}\equiv \mbox{arccosh} (2/\Omega_{\rm f} -1)$, $w_{\rm
  f}\equiv 1/\Omega_{\rm f} - 1$, $\chi \equiv \lambda_0 H_0^2 r_{\rm
  ta}^3/(GM)$, $r_{\rm ta}$ is the maximum turn-around radius and $F$
is the hyper-geometric function of type (2,1). The density parameter
$\Omega_{\rm f}$ is defined at $z_{\rm f}$ and is related to
$\Omega_0$ and $\lambda_0$ through
\begin{eqnarray}
\Omega_{\rm f} &=& {\Omega_0(1+z_{\rm f})^3 \over
\Omega_0(1+z_{\rm f})^3+(1-\Omega_0-\lambda_0)(1+z_{\rm f})^2+\lambda_0} . 
\end{eqnarray}

\section{\normalsize\bf Mass variance for the CDM fluctuation spectrum}
\setcounter{equation}{0}

The mass variance $\sigma^2(M)$ is related to the linear power
spectrum of density fluctuations $P(k)$ through
\begin{equation}
  \sigma^2(M) = {1 \over (2\pi)^3} \int_0^\infty P(k) W^2(k r_\M)4\pi
  k^2 dk,  
\label{eq:variance}
\end{equation}
where the filtering radius $r_\M$ is related to the mass $M$ by $r_\M
= [3M/(4\pi \rho_\0)]^{1/3}$ as the top-hap window function is used: 
\begin{equation}
  W(x) = \frac{3}{x^3}(\sin x - x \cos x).
\end{equation}
We adopt the CDM power spectrum of the form given by Bardeen et
al. (1986) with the scale invariant initial condition:
\begin{equation}
  P(k) \propto k \lbkt{\ln(1+2.34q) \over 2.34 q}^2 \lbkt{1 + 3.89 q +
    (16.1q)^2 + (5.46q)^3 + (6.71q)^4}^{-1/2},
\label{eq:cdmspect}
\end{equation}
where $q\equiv k/(\Gamma h \mbox{ Mpc}^{-1})$. Here the quantity
$\Gamma$ for non-negligible baryon density is given by (Peacock \&
Dodds 1994; Sugiyama 1995)
\begin{equation}
  \Gamma = \Omega_0 h (T_0/2.7 \mbox{ K})^{-2}\exp [-\Omega_{\rm
    B}(1+\sqrt{2 h}\Omega_0^{-1})],
\end{equation}
where $T_0$ is the temperature of the cosmic microwave background
radiation. 

We found that $\sigma^2$ and $d\sigma^2/dM$ calculated from the above
equations are accurately fitted simultaneously by the following
formula and its derivative with respect to $m$:
\begin{equation}
  \sigma^2 \propto \lbkt{1 + 2.208 m^p - 0.7668 m^{2p} +
    0.7949 m^{3p}}^{-{2 \over 9p}},
\label{eq:cdmfit}
\end{equation}
where $p=0.0873$, and $m\equiv M(\Gamma h)^2/(10^{12}\msun)$.  The
above approximation holds within a few percent for both $\sigma^2$ and
$d\sigma^2/dM$ in the range $10^{-6} \nlt m \nlt 10^{4}$. Throughout
the present paper, $\sigma$ is evaluated via equation
(\ref{eq:cdmfit}) and normalized by
\begin{equation}
\sigma(r_{\rm M}= 8h^{-1}\mbox{Mpc}) = \frac{1}{b}, 
\label{eq:bias}
\end{equation}
where $b$ is the bias parameter. 

\section{\normalsize\bf Fitting formulae for the Lacey \& Cole 
distribution function}
\setcounter{equation}{0}

Since the distribution function of the halo formation epochs derived
by LC (eq.~[\ref{eq:dpdwf}]) requires a time-consuming numerical
integration, the following fitting formula is used in the present
paper.

First we consider the case of power-law fluctuation spectra $P(k)
\propto k^n$, i.e. $\sigma^2(M)\propto M^{-\alpha}$ where
$\alpha\equiv (n+3)/3$. Then equation (\ref{eq:dpdwf}) reduces to
\begin{eqnarray} 
\label{eq:dpdwalpha} 
{\partial p \over \partial \tilde \omega_{\rm f}}
(\alpha, \tilde\omega_{\rm f}) 
&=& \sqrt{8\over \pi} {\rm e}^{- \tilde \omega_{\rm f}^2/2} -
{2(2^\alpha-1) \over \sqrt{\pi}\alpha} \tilde \omega_{\rm f}
\int_{\tilde \omega_{\rm f} \over \sqrt{2}}^\infty dy { {\rm e}^{-y^2}
  \over y^2 } \left( 1+ {2^\alpha-1 \over 2}{\tilde \omega_{\rm f}^2
    \over y^2} \right)^{(1-\alpha)/\alpha} .
\end{eqnarray}
For $\tilde \omega_{\rm f} \ll 1$, equation (\ref{eq:dpdwalpha}) becomes
\begin{eqnarray}
\label{eq:dpdwflinear}
{\partial p \over \partial \tilde \omega_{\rm f}} (\alpha,
\tilde\omega_{\rm f}) &\simeq& \sqrt{8 \over \pi} \left[ 1 - {
    \sqrt{2^\alpha-1} \over \alpha } \int_0^{\sqrt{2^\alpha-1}}
  (1+x^2)^{(1-\alpha)/\alpha} dx \right] \nonumber \\ && + {
  2(2^\alpha-1) \over \alpha }\tilde \omega_{\rm f} + O(\tilde
\omega_{\rm f}^2) ,
\end{eqnarray}
For $\tilde \omega_{\rm f} \gg 1$, on the other hand, we found that
numerical results for equation (\ref{eq:dpdwalpha}) are insensitive to
$\alpha$ and well approximated by the solution in the $\alpha=1$ case:
\begin{eqnarray}
\label{eq:dpdwf1}
{\partial p \over \partial \tilde \omega_{\rm f}} (\alpha=1,
\tilde\omega_{\rm f}) = 2 \tilde \omega_{\rm f} \, {\rm erfc}\left(
  {\tilde \omega_{\rm f} \over \sqrt{2}} \right) .
\end{eqnarray}

On the basis of equations (\ref{eq:dpdwflinear}) and
(\ref{eq:dpdwf1}), we constructed the following interpolation formula
which gives a very good fit to equation (\ref{eq:dpdwalpha}):
\begin{eqnarray}
\label{eq:dpdwfit}
{\partial p \over \partial \tilde \omega_{\rm f}} (\alpha,
\tilde\omega_{\rm f}) &=& { A(\alpha) \over 1 + B(\alpha)\tilde
  \omega_{\rm f}} {\rm e}^{-5\tilde \omega_{\rm f}^2} + 2 C(\alpha)
\tilde \omega_{\rm f} \, {\rm erfc}\left( {\tilde \omega_{\rm f} \over
    \sqrt{2}} \right) ,
\end{eqnarray}
where
\begin{eqnarray}
  A(\alpha) &\equiv& \sqrt{8 \over \pi}(1-\alpha)(0.0107 +
  0.0163\alpha), \\ 
  B(\alpha) &\equiv& {2 \over A(\alpha)} \left[ C(\alpha) -
    {2^\alpha-1 \over \alpha} \right] , \\ 
  C(\alpha) &\equiv& 1 - {1-\alpha \over 25}.
\end{eqnarray}

\begin{figure}
\begin{center}
   \leavevmode\psfig{figure=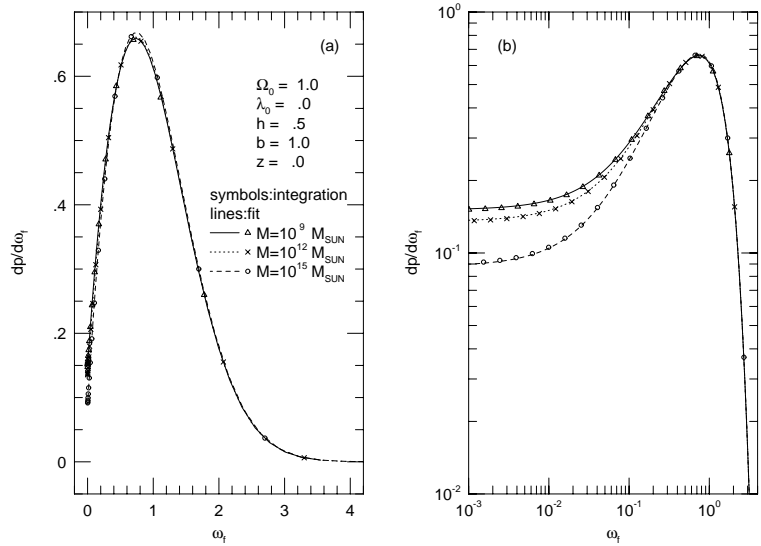,height=8cm,angle=90}
\end{center}
\caption{The scaled distribution function of halo formation epochs, 
  $\partial p/\partial \tilde \omega_{\rm f}$, in the standard CDM
  model ($\Omega_0=1$, $\lambda_0=0$ and $h=0.5$); (a) linear and (b)
  log-log plots. Symbols and lines indicate results of a numerical
  integration of equation (\protect\ref{eq:dpdwf}\protect) and a fit
  by equations
  (\protect\ref{eq:dpdwfit}\protect)--(\protect\ref{eq:alphaeff}\protect),
  respectively; $M=10^9 \msun$ (triangles, solid line), $10^{12}
  \msun$ (crosses, dotted line), and $10^{15} \msun$ (circles, dashed
  line).}
\label{fig:dpdw}
\end{figure}

Since equations (\ref{eq:dpdwalpha})--(\ref{eq:dpdwf1}) are valid only
when the underlying spectra are of power-law form, the above fitting
formulae may not be applicable to more general spectra.  In fact, in
the case of the CDM spectrum, a straightforward substitution of the
effective spectral index $\alpha_{\rm eff} \equiv - d\ln\sigma_{\rm
  CDM}^2/d\ln M$ into equation (\ref{eq:dpdwfit}) does not work very
well, because $\alpha_{\rm eff}$ varies with the mass scale and so
cannot be approximated by a constant in the integral of equation
(\ref{eq:dpdwf}).  Nevertheless, we found that equation
(\ref{eq:dpdwfit}) still gives an excellent empirical fit to equation
(\ref{eq:dpdwf}) for the CDM spectrum, if $\tilde \alpha_{\rm eff}$
below is adopted rather than $\alpha_{\rm eff}$:
\begin{equation}
\label{eq:alphaeff}
\tilde \alpha_{\rm eff} \equiv \alpha_{\rm eff}(0.6268 + 0.3058
\alpha_{\rm eff}).
\end{equation}
The accuracy of this approximation is within a few percent in the
range $10^6\msun \nlt M(\Omega_0 h^2)^2 \nlt 10^{16}\msun$ (see
Fig.~\ref{fig:dpdw}).

\vfill\eject \bigskip

\bigskip
\centerline{\bf REFERENCES}
\bigskip

\def\apjpap#1;#2;#3;#4; {\pp#1, {#2}, {#3}, #4}
\def\apjbook#1;#2;#3;#4; {\pp#1, {#2} (#3: #4)}
\def\apjppt#1;#2; {\pp#1, #2.}
\def\apjproc#1;#2;#3;#4;#5;#6; {\pp#1, {#2} #3, (#4: #5), #6}

\apjpap Abramopoulos, F., \& Ku, W. 1983; ApJ; 271; 446;   
\apjpap Bardeen, J.M., Bond J.R., Kaiser, N., \& Szalay, A.S. 1986;
ApJ; 304; 15; 
\apjpap Blain, A.W., \& Longair, M.S. 1993; MNRAS; 265; L21;   
\apjpap Bond, J.R., Cole, S., Efstathiou, G., \& Kaiser, N. 1991; ApJ; 
379; 440;
\apjpap Bower, R.J. 1991; MNRAS; 248; 332;  
\apjpap Bryan, G.L., Cen, R., Norman, M.L., Ostriker, J.P., \& Stone,
J.M. 1994; ApJ; 428; 405;
\apjpap Cen, R., Kang, H., Ostriker, J.P., \& Ryu, D. 1995; ApJ; 451; 436;
\apjpap David, L.P., Slyz, A., Jones, C., Forman, W., \& Vrtilek, 
S.D. 1993; ApJ; 412; 479;  
\apjpap Edge, A.C., \& Stewart, G.C. 1991a; MNRAS; 252; 414;
\apjpap Edge, A.C., \& Stewart, G.C. 1991b; MNRAS; 252; 428;
\apjpap Evrard, A.E. 1990; ApJ; 363; 349;  
\apjpap Evrard, A.E., \& Henry, J.P. 1991; ApJ; 383; 95;
\apjpap Gioia, I.M., Henry, J.P., Maccacaro, T., Morris, S.L., Stocke,
J.T., \& Wolter, A. 1990; ApJ; 356; L35;
\apjpap Gorski, K.M., Ratra, B., Sugiyama, N., \& Banday, A.J. 1995;
ApJ; 444; L65;   
\apjpap Hanami, H. 1993; ApJ; 415; 42; 
\apjpap Henry, J.P., \& Arnaud, K.A. 1991; ApJ; 372; 410; 
\apjpap Jones, C., \& Forman, W. 1984; ApJ; 276; 38;
\apjpap Kaiser, N. 1991; ApJ; 383; 104; 
\apjpap Kang, H., Cen, R., Ostriker, J.P., \& Ryu, D. 1994; ApJ; 428; 1;
\apjppt Kitayama, T., \& Suto, Y. 1996; MNRAS, in press (KS);
\apjpap Lacey, C.G., \& Cole, S. 1993; MNRAS; 262; 627 (LC);
\apjpap Lahav, O., Lilje, P.B., Primack, J.R., \& Rees, M.J. 1991;
MNRAS; 251; 128;
\apjppt Nakamura, T.T. 1996; Master thesis (Univ. of Tokyo, unpublished);
\apjpap Navarro, J.F., Frenk, C.S., \& White, S.D.M. 1995; MNRAS; 275; 720;
\apjpap Peacock, J.A., \& Dodds, S.J. 1994; MNRAS; 267; 1020;
\apjbook Peebles, P.J.E. 1980; The Large-Scale Structure of the
  Universe.; Princeton Univ. Press; Princeton;
\apjpap Press, W.H., \& Schechter, P. 1974; ApJ; 187; 425;
\apjbook Sarazin, C.L. 1988; X-ray Emission from Clusters of
 Galaxies.; Cambridge Univ. Press; Cambridge;
\apjpap Sasaki, S. 1994; PASJ; 46; 427; 
\apjppt Sasaki, S., Masai, K., Kitayama, T., \& Suto, Y. 1996; in preparation; 
\apjpap Suginohara, T. 1994; PASJ; 46; 441;
\apjpap Sugiyama, N. 1995; ApJS; 100; 281; 
\apjpap Suto, Y. 1993; Prog.Theor.Phys.; 90; 1173;
\apjpap Suto, Y., Makishima, K., Ishisaki, Y., \& Ogasaka, Y. 1996;
ApJL; 461; L33;
\apjpap Tanvir, N.R., Shanks, T., Ferguson, H.C., \& Robinson,
D.R.T. 1995; Nature; 377; 27;
\apjpap Walker, T.P., et al. 1991; ApJ; 376; 51; 

\vfill\eject

\newpage

\end{document}